\documentclass[reprint,pre,showpacs,amsmath,amssymb,aps]{revtex4-1}

\usepackage{natbib}
\usepackage{graphicx,color,rotating}
\usepackage[latin1]{inputenc}
\usepackage{textcomp}
\usepackage{dcolumn}
\usepackage{bm}     
\usepackage[version=3]{mhchem}  
\usepackage{upgreek}

\newcommand{\notop}{{{}_{}}}

\newcommand{\mr}[1]{\ensuremath{\mathrm{#1}}}
\renewcommand{\vec}[1]{\bm{#1}}

\newcommand{\pp}{\partial}       

\newcommand{\nablabf}{\boldsymbol{\nabla}}

\newcommand{\vJp}{\vec{J}_+}
\newcommand{\vJm}{\vec{J}_-}
\newcommand{\vJpm}{\vec{J}_{\pm}}

\newcommand{\vn}{\vec{n}}

\newcommand{\cpm}{c_\pm}

\newcommand{\Dpm}{D^\notop_{\pm}}

\newcommand{\kB}{{k^\notop_\mathrm{B}}}

\newcommand{\eps}{\epsilon}
\newcommand{\epsw}{\epsilon_{\mathrm{w}}}

\newcommand{\bgam}{\bar{\gamma}}

\newcommand{\kmax}{k_\mr{max}}

\newcommand{\lamD}{\lambda^{{}}_\mathrm{D}}
\newcommand{\blam}{\bar{\lambda}}
\newcommand{\blamD}{\bar{\lambda}^{{}}_\mathrm{D}}

\newcommand{\blamDsqr}{\bar{\lambda}^{{2}}_\mathrm{D}}

\newcommand{\VT}{V_{\mathrm{T}}}
\newcommand{\mupm}{{\mu_{\pm}^{{}}{}}}

\newcommand{\SIC}{\textrm{C}}

\newcommand{\SIF}{\textrm{F}}

\newcommand{\SIJ}{\textrm{J}}

\newcommand{\SIK}{\textrm{K}}

\newcommand{\SIm}{\textrm{m}}

\newcommand{\SImM}{\textrm{mM}}

\newcommand{\SImm}{\textrm{mm}}
\newcommand{\SImum}{\textrm{\textmu{}m}}
\newcommand{\SInm}{\textrm{nm}}

\newcommand{\SIs}{\textrm{s}}

\newcommand{\beq}[1]{\begin{equation} \eqlab{#1}}
\newcommand{\eeq}{\end{equation}}
\newcommand{\bsub}{\begin{subequations}}
\newcommand{\esub}{\end{subequations}}
\def\bal#1\eal{\begin{align}#1\end{align}}
\def\bsubal#1\esubal{\bsub \begin{align}#1\end{align} \esub}
\newcommand{\nn}{\nonumber}
\newcommand{\eqlab}[1]{\label{eq:#1}}
\renewcommand{\eqref}[1]{Eq.~(\ref{eq:#1})}
\newcommand{\eqrefNoeq}[1]{(\ref{eq:#1})}
\newcommand{\eqsref}[2]{Eqs.~(\ref{eq:#1}) and~(\ref{eq:#2})}

\newcommand{\figref}[1]{Fig.~\ref{fig:#1}}

\newcommand{\figlab}[1]{\label{fig:#1}}

\newcommand{\appsref}[2]{Appendices~\ref{chap:#1} and~\ref{chap:#2}}

\newcommand{\chaplab}[1]{\label{chap:#1}}

\newcommand{\tabref}[1]{Table~\ref{tab:#1}}

\newcommand{\tablab}[1]{\label{tab:#1}}

\begin{document}

\title{Morphological instability during steady electrodeposition at overlimiting currents}	

\author{Christoffer P. Nielsen}
\affiliation{Department of Physics, Technical University of Denmark, DTU Physics Building 309, DK-2800 Kongens Lyngby, Denmark}
\email{chnie@fysik.dtu.dk and bruus@fysik.dtu.dk}
\author{Henrik Bruus}
\affiliation{Department of Physics, Technical University of Denmark, DTU Physics Building 309, DK-2800 Kongens Lyngby, Denmark}

\date{27 May 2015, submitted to Phys.\ Rev.\ E}

\begin{abstract}

We present a linear stability analysis of a planar metal electrode during steady electrodeposition. We extend the previous work of Sundstrom and Bark by accounting for the extended space-charge density, which develops at the cathode once the applied voltage exceeds a few thermal voltages. In accordance with Chazalviel's conjecture, the extended space-charge region is found to greatly affect the morphological stability of the electrode. To supplement the numerical solution of the stability problem, we have derived analytical expressions valid in limit of low and high voltage, respectively.

\end{abstract}
\maketitle


\section{Introduction}

One of the most interesting aspects of systems, involving transport between matter in different phases, is their tendency to become morphologically unstable and develop ramified growth patterns. Well known examples include snow flake formation and dendritic growth during metal solidification \cite{Libbrecht2005,Trivedi1994}. A particularly interesting and challenging growth problem is encountered in electrodeposition from an electrolyte onto an electrode \cite{Brady1984, Nikolic2006, Kahanda1989, Leger2000, Gonzalez2008, Han2014, Trigueros1991, Devos2007, Nishikawa2013}. Whereas snow flake formation and solidification are mainly driven by diffusion of water vapour and heat, respectively \cite{Libbrecht2005, Trivedi1994}, electrodeposition is driven by electromigration in addition to diffusion \cite{Chazalviel1990, Rosso2007}. For this reason, the electrodeposition rate can be driven to exceed the diffusion limit, at which point the system enters a nonlinear regime not encountered in the purely diffusion-driven systems. One of the features of this nonlinear regime is the development of a nonequilibrium space-charge region, which extends from the cathode into the electrolyte \cite{Chazalviel1990, Rosso2006, Nielsen2014}. Already in 1990, Chazalviel realized that this extended space-charge region is crucial to the understanding of ramified growth during electrodeposition \cite{Chazalviel1990}. Nevertheless, there has been very little work which actually takes this effect into account.

In this paper we investigate the morphological stability of the cathode during electrodeposition in both the linear and the nonlinear regime. We follow the approach of Sundstrom and Bark \cite{Sundstrom1995}, and investigate steady electrodeposition in a system composed of an electrolyte sandwiched between two, initially planar, metal electrodes. We solve the stability problem numerically and find that the higher the applied voltage difference is, the more unstable the electrode surface becomes. Also, the most unstable wavelength becomes smaller as the voltage bias is increased.

In addition to solving the stability problem numerically, we derive analytical expressions for the perturbation growth rate, valid in the low and high voltage limit, respectively. In deriving these expressions, we make use of an accurate analytical model for the extended space-charge region, which we presented in a recent paper \cite{Nielsen2014}.

\section{Model system}

Following Sundstrom and Bark \cite{Sundstrom1995}, we consider a binary electrolyte trapped between two co-planar metal electrodes at $x=0$ and $x=2L$. The electrolyte has initial concentration $c_0$ and is assumed symmetric with valence $Z$. The coordinate system is moving in the negative $x$-direction with velocity $U$, which is related to the mean movement of the electrode. We consider the dilute solution limit, in which the effect of the moving coordinate system is negligible everywhere except in the surface evolution equation.
A sketch of the system is shown in \figref{My_Geometry}.

In the analysis, we investigate the stability of $y$-dependent perturbations along the $x$-direction. However, our analysis is general and applies to perturbations along any direction in the $yz$ plane.

\begin{figure}[!t]
    \includegraphics[width=0.9\columnwidth]{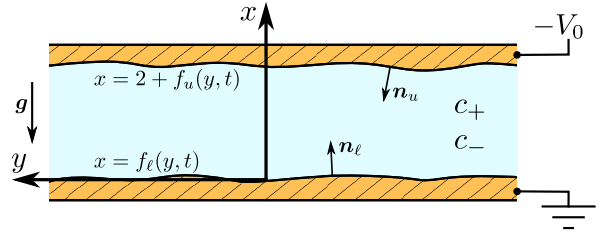}
    \caption{\figlab{My_Geometry} Sketch of the studied system with lower and upper electrode surfaces at $x=f_\ell(y,t)$ and $x=2+f_u(y,t)$, respectively. The coordinates are given relative to the moving frame of reference, following the mean speed of the electrode surfaces, and normalized by half the electrode spacing $L$.}
\end{figure}

\section{Governing equations}

The current densities of either ion are given as
 \bsubal
 2\vJpm &= - \cpm \nablabf \mupm, \\
 \mupm &= \ln(\cpm) \pm Z\phi,
 \esubal
where we have non-dimensionalized the currents $\vJpm$ by the limiting currents $2\Dpm c_0/L$, the electrochemical potentials $\mupm$ by $\kB T$, the electric potential $\phi$ by the thermal voltage $\VT = \kB T/e$, the coordinates by half the electrode spacing $L$, and the concentrations $\cpm$ by the initial concentration $c_0$. Normalizing the time by the diffusion time $t_0 = L^2/(2D_+)$, the non-dimensionalized ion-conservation equations become
\begin{align}
\frac{D_+}{\Dpm}\pp_t \cpm &= - \nablabf \cdot \vJpm.
 \end{align}
At the electrodes, the current of anions vanishes, while the current of cations is given by a reaction expression
 \bsub
 \eqlab{currentBC}
 \begin{align}
 \vn_p \cdot \vJm &=0, \\
 \vn_p \cdot \vJp &=-R_p, \eqlab{JpBC}
 \end{align}
 \esub
where $R_p$ is the reaction rate at the lower and upper electrode, respectively, as indicated by the subscript $p = \ell, u$.
We model the reaction rates $R_u$ and $R_\ell$ using the standard Butler-Volmer expression \cite{Sundstrom1995},
\begin{align}  \eqlab{Butler_Volmer}
R_p=K_0  \left [ c_+ e^{-\bgam \kappa+\alpha Z(\phi+V_p)}-e^{-\bgam \kappa -   (1-\alpha)Z(\phi+V_p) }\right ],
\end{align}
where $K_0$ is the dimensionless version of the dimensionfull rate constant $k_0$ for the electrode reaction,
\begin{align}
K_0 = \frac{k_0}{2D_+c_0/L}.
\end{align}
Above, $V_p$ is the normalized electrode potential, $\kappa$ is the normalized curvature of the surface, $\alpha$ is the charge-transfer coefficient, and $\bgam$ is the non-dimensionalized version of the dimensionfull surface energy $\gamma$,
\begin{align}
\bgam = \frac{a^3 \gamma}{\kB T L}.
\end{align}

The electrostatic part of the problem is governed by the Poisson equation,
\begin{align}
2 \blamDsqr \nabla^2 \phi  = -\rho =  -Zc_++Zc_-,
\end{align}
where the non-dimensional Debye length $\blamD$ is given as
\begin{align}
\blamD = \frac{\lamD}{L},\quad  \text{with} \quad   \lamD = \sqrt{\frac{\kB T \epsw}{2e^2c_0}}.
\end{align}
For simplicity, and to be in accordance with most previous work, we choose not to explicitly model the Debye layers adjoining the electrodes. Instead, we apply the boundary conditions \eqrefNoeq{currentBC} just outside the Debye layer.
Following Ref.~\cite{Nielsen2014b} we implement the boundary condition
\begin{align}  \eqlab{nablacpBC}
\vn_u \cdot \nablabf c_+ = 0,
\end{align}
at the upper electrode, to reflect the minimum in $c_+$ at the outer edge of the Debye layer. Together with \eqref{JpBC}, condition \eqrefNoeq{nablacpBC} corresponds to ascribing the entire current into the upper electrode to electromigration.

Finally, since the anions can not enter or leave the system the total number of anions is conserved,
\begin{align}
\int_\Omega \big(c_- - 1\big) \ \mr{d}V=0.
\end{align}

We introduce functions $x=f_p(y)$ describing the position of the upper and lower electrode $u$ and $\ell$. The time evolution of $f_p$ is determined by the single-ion volume $a^3$ and the current into the electrode,
\bsub \eqlab{Electrode_growth}
\begin{align}
\left(\pp_t f_\ell - U \right )\vec{e}_x \cdot \vn_\ell&= -a^3c_0\vn_\ell \cdot \vJp,  & &\text{Anode}, \\
\left(\pp_t f_u - U \right )\vec{e}_x \cdot \vn_u&= -a^3c_0\vn_u \cdot \vJp, & & \text{Cathode.}
\end{align}
\esub
Here, the filling factor $a^3c_0$ is much less than unity, since we are dealing with dilute solutions.  The normalized velocity $U$ of the coordinate system accounts for the mean current into or out off the electrodes, and $\pp_t f_p$ accounts for local deviations from the mean current.

The curvature $\kappa$ and the normal vectors are related to the surface function $f_p$ by
\bsub \eqlab{Full_normal_and_curv}
\begin{align}
\vn_\ell &= \frac{\vec{e}_x-\vec{e}_y\pp_y f_\ell}{\sqrt{1+(\pp_y f_\ell)^2}}, \quad \vn_u = \frac{-\vec{e}_x+\vec{e}_y\pp_y f_u}{\sqrt{1+(\pp_y f_u)^2}}, \\
\kappa_\ell &= \frac{\pp_y^2 f_\ell}{\sqrt{1+(\pp_y f_\ell)^2}}, \quad \kappa_u = -\frac{\pp_y^2 f_u}{\sqrt{1+(\pp_y f_u)^2}}.
\end{align}
\esub

In defining the above equations and boundary conditions, we have chosen slightly different normalizations than in Ref.~\cite{Sundstrom1995}, the main difference being that we allow for a non-zero space charge density.

\section{Perturbation}

The stability of the problem is investigated using linear perturbation theory. That is, we impose a small perturbation on a steady-state base state, and investigate how the perturbation evolves. The base state is identified by a superscript $"0"$ and the first-order perturbation by superscript $"1"$,
 \bsubal
 f_p(y,t) &\approx  f_p^1(y,t), \\
 \cpm(x,y,t) &\approx \cpm^0(x) +  \cpm^1(x,y,t) , \\
 \phi(x,y,t) &\approx \phi^0(x) +  \phi^1(x,y,t) .
 \esubal
In first-order perturbation theory, we substitute the second-order factor $\sqrt{1+(\pp_y f_p)^2}$ in \eqref{Full_normal_and_curv} by unity,
\bsub \eqlab{Approx_normal_and_curv}
\begin{align}
\vn_\ell &\approx \vec{e}_x -\vec{e}_y\pp_y f^1_\ell, &\vn_u &\approx  -\vec{e}_x+\vec{e}_y \pp_y f^1_u, \\
\kappa_\ell &\approx  \pp_y^2 f^1_\ell, &\kappa_u &\approx -\pp_y^2 f^1_u.
\end{align}
\esub
To find the field values at the perturbed surface, we Taylor expand to first order and obtain
 \bsubal
 \phi (  f_\ell^1,y,t) & \approx \phi^0(0)+\pp_x\phi|_0  f_\ell^1(y,t) +  \phi^1(0,y,t), \\
 \nablabf \phi( f_\ell^1,y,t) &\approx \eps \pp_y \phi^1|_0 \vec{e}_y  \\
 &\quad +\left ( \pp_x \phi^0|_0 + \pp_x^2 \phi^0|_0\eps f^1_\ell + \eps \pp_x \phi^1|_0\right )\vec{e}_x. \nn
 \esubal
Similar expressions apply for $\cpm$ and at the upper electrode.
Evaluating the reaction rate at the lower electrode and expanding to first order, we find
 \bsubal
  R_\ell &\approx R_\ell^0 +  R_\ell^1,
 \\
 \eqlab{Zero_order_reaction}
 \frac{R_\ell^0}{K_0} &=  c_+^0 e^{\alpha Z(\phi^0+V_\ell)}-
 e^{-(1-\alpha)Z(\phi^0+V_\ell) },
 \\
 \frac{R_\ell^1}{K_0} &=  e^{\alpha Z(\phi^0+V_\ell)}\bigg[ c_+^1+ \pp_x c^0 f_\ell^1
 \\ \nn
 &\quad +c_+^0\Big(-\bgam \pp_y^2 f_\ell^1+\alpha Z\big[\phi^1+\pp_x \phi^0 f_\ell^1\big]\Big) \bigg]
 \\ \nn
 &\quad -e^{ - (1-\alpha)Z(\phi^0+V_\ell) }\bigg[-\bgam \pp_y^2 f_\ell^1
 \\ \nn
 &\quad -   (1-\alpha)Z\big[\phi^1+\pp_x \phi^0 f^1_\ell \big] \bigg],
 \esubal
where all fields are evaluated at $x=0$. Similar expressions apply at the upper electrode.

Hence, the full zeroth-order problem becomes
 \bsubal
 0 &= - \pp_x  J_\pm^0, \\
 2J_\pm^0 &= -\pp_x \cpm^0 \mp Z\cpm^0\pp_x \phi^0, \\
 2 \blamDsqr \pp_x^2 \phi^0  &=   -Z(c_+^0-c_-^0) = -\rho^0,
 \esubal
with the following boundary conditions and constraints
 \bsubal
 J_-^0(0) &= 0, & J_-^0(2)&=0, \\
 J_+^0(0) &= -R_\ell^0, & J_+^0(2) &= R_u^0, \\
\int_0^2 \big( c_-^0-1 \big)\ \mr{d}x&=0, &  \pp_x c_+^0(2) &=0,
 \esubal
and the mean growth velocity $U$ derived from \eqref{Electrode_growth},
\begin{align}
U = a^3 c_0 J_+^0.   \eqlab{zero_order_electrode_growth}
\end{align}

Similarly, the first-order problem is given by
 \bsubal
 \frac{D_+}{\Dpm}\pp_t \cpm^1 &= - \nablabf \cdot  \vJpm^1, \\
 2\vJpm^1 &= -\nablabf \cpm^1 \mp Z\cpm^0\nablabf \phi^1\mp Z\cpm^1\nablabf \phi^0, \\
 2 \blamDsqr \nabla^2 \phi^1  &=   -Z(c_+^1-Zc_-^1),
 \esubal
and the boundary conditions,
 \bsubal
\vec{e}_x \cdot  \vJm^1(2)&=0, &  \vec{e}_x \cdot  \vJm^1(0)&=0,\\
 \vec{e}_x \cdot  \vJp^1(2)&=R_u^1, & \vec{e}_x \cdot  \vJp^1(0)&=-R_\ell^1,\\
 \pp_x^2 c_+^0(2) f_u^1 + \pp_x c_+^1(2) &=0, & &
 \esubal
together with the first-order electrode growth rates $\pp_t f^1_\ell$ and $\pp_t f^1_u$ derived from \eqref{Electrode_growth},
 \begin{align} \eqlab{First_order_growth}
 \pp_t f^1_\ell &= a^3c_0R_\ell^1, & \pp_t f^1_u &=- a^3c_0R_u^1.
 \end{align}

To find the eigenmodes, we make the following harmonic ansatz for the first-order fields,
 \bsub
 \begin{align}
 \cpm^1(x,y,t) &= \cpm^*(x) e^{\Gamma t + iky}, \\
 \phi^1(x,y,t) &= \phi^*(x)e^{\Gamma t + iky}, \\
 f_p^1(y,t) &= F_pe^{\Gamma t + iky},
 \end{align}
where $\Gamma$ is the nondimensional growth rate of the perturbation, and $k$ is the wavenumber of the transverse eigenmode. For convenience we also define
 \begin{align}
 R^1_p = R^*_pe^{\Gamma t + iky}.
 \end{align}
 \esub
With this ansatz, the first-order bulk equations become
 \bsubal
 2\frac{D_+}{D_\pm} \Gamma\cpm^* &=
 - k^2 (\cpm^* \pm Z\cpm^0\phi^*) \eqlab{Star_bulk_eq}
 \\ \nn
 & +\pp_x \Big\{\pp_x \cpm^* \pm Z \cpm^*\pp_x\phi^0\pm Z \cpm^0\pp_x \phi^*  \Big\}
 \\
 2\blamDsqr (\pp_x^2 \phi^* - k^2 \phi^* ) &= -Z(c_+^* -c_-^*), \eqlab{Star_Poisson}
 \esubal
and the first-order reaction rate at the lower electrode is
 \begin{align}
 \frac{R_\ell^*}{K_0} &=  e^{\alpha Z(\phi^0+V_\ell)}\bigg[ c_+^*+ \pp_x c^0 F_\ell
 \nn \\
 &\quad +c_+^0 \Big(-\bgam k^2F_\ell+\alpha Z\big[\phi^*+\pp_x \phi^0 F_\ell\big]\Big) \bigg]
 \nn \\
 &\quad -e^{ - (1-\alpha)Z(\phi^0+V_\ell) }\bigg[-\bgam k^2 F_\ell
 \nn \\
 &\quad -   (1-\alpha)Z\big[\phi^*+\pp_x \phi^0 F_\ell\big] \bigg].
\end{align}
Inserting the ansatz in the growth equations~\eqrefNoeq{First_order_growth} yields
\begin{align}
\Gamma F_\ell &= a^3c_0 R_\ell^*, & \Gamma F_u &= -a^3c_0 R_u^*.
\end{align}

\section{Analytical results}

For large wavenumbers, $k\gtrsim 1$, we can neglect $f_\ell$ and the left hand side in \eqref{Star_bulk_eq}. Analytical expressions for the growth rate can then be obtained in the limit of overlimiting and underlimiting current, respectively. In \appsref{App_Lin}{App_Nonlin} we find that the growth rate can be expressed as
 \begin{align}  \eqlab{bGam_analyt}
 \Gamma = a^3c_0kJ^0 \frac{\xi - \bgam k^2}{\xi + k}.
 \end{align}
With the usual assumption $\alpha=\frac12$, simple expressions for $\xi$ can be obtained. In the limit of underlimiting current $J^0<1$ it becomes
 \begin{align}
 \xi =\frac{J^0}{1-J^0}\left [\frac{1}{2} +\sqrt{1+4\left (\frac{K_0}{J^0}\right )^2(1-J^0)}\right ].
 \end{align}
We note that the underlimiting expression is nearly identical to the one already derived by Sundstrom and Bark \cite{Sundstrom1995}. In the limit of overlimiting current $J^0>1$, we obtain
 \bsubal
 \xi &=\frac{J^0}{c_+^0}\sqrt{1+4\left (\frac{K_0}{J^0}\right )^2c_+^0}, \\
 c_+^0 &\approx \frac{\blamD}{Z} \sqrt{\frac{2J^0}{1-\frac{1}{J^0}}}.
 \esubal
The critical wavenumber $k_\mr{c}$, where the perturbation is marginally stable, is found to be
 \bsub
 \begin{align}
 k_\mr{c} = \sqrt{ \frac{\xi}{\bgam}},
 \end{align}
and the wavenumber $\kmax$, at which the growth rate is maximum, is given as
 \begin{align}
 \kmax &= \frac{\xi}{2}\bigg [ \left (\frac{2-\xi\bgam+2\sqrt{1-\xi\bgam}}{\xi\bgam}\right )^{1/3} \nn \\
 &\quad +\left (\frac{2-\xi\bgam+2\sqrt{1-\xi\bgam}}{\xi\bgam}\right )^{-1/3}  -1 \bigg].
 \end{align}
 \esub

We note that the analytical model takes the zeroth-order current density $J^0$ as input variable through $\xi$. If one wants the results as a function of the potential drop instead, a model of the system's current-voltage characteristic is needed. For simplicity, we just use the numerically calculated current-voltage characteristic in the following.

 To compute the results without reference to a numerical solution, an analytical model for the system's current-voltage characteristic is required. Such a model can be found in our previous work \cite{Nielsen2014}. To obtain the total voltage drop over the system, the interfacial voltages from \eqref{Zero_order_reaction} should also be taken into account.

\begin{table}[!b]
\caption{\tablab{Parameters} Fixed parameter values used in the numerics.}
\begin{ruledtabular}
\begin{tabular}{lcl}
Parameter & Symbol & Value  \\ \hline

Cation diffusivity\cite{Lide2010} & $D_+$ & $0.714\times 10^{-9}\SIm^2\: \SIs^{-1}$ \rule{0mm}{2.5ex} \\

Anion diffusivity\cite{Lide2010} & $D_-$ & $1.065\times 10^{-9}\SIm^2 \: \SIs^{-1}$ \\

Ion valence & $Z$ &  $2$ \\

Surface energy & $\gamma$ &$1.85\ \SIJ\: \SIm^{-2}$ \\

Temperature & $T$ & $300\ \SIK$ \\

Permittivity of water & $\epsw$ & $6.90\times 10^{-10}\SIF\: \SIm^{-1}$ \\

Charge-transfer coefficient & $\alpha$ & $\frac12$ \\

Reaction constant\footnote{Calculated using the exchange current $I_0=30\ \mr{A}\: \SIm^{-2}$ from Ref.~\cite{Turner1962} and $k_0 = I_0/(Ze)$.} & $k_0$ & $9.4\times 10^{19} \SIm^{-2}\: \SIs^{-1}$\\

Diameter of a copper atom\footnote{The cube root of the volume per atom in solid copper \cite{Lide2010}.} & $a$ & $0.228\ \SInm$ \\

\end{tabular}
\end{ruledtabular}
\end{table}

\section{Numerical solution}

The numerical simulations are carried out in the commercially available finite element software \textsc{COMSOL Multiphysics} ver.~4.3a. Following our previous work \cite{Gregersen2009, Nielsen2014, Nielsen2014b}, the zeroth- and first-order problems are rewritten in weak form and implemented in the mathematics module of \textsc{COMSOL}. In the first-order problem we set the parameter $f_u$ to unity, meaning that the magnitude of the remaining first-order fields are given relative to the amplitude of the upper electrode perturbation. To limit the parameter space, we choose fixed, physically reasonable values for the parameters listed in \tabref{Parameters}. The values are chosen to correspond to copper electrodes in a copper sulfate solution. We note that the surface tension is quite difficult to determine experimentally, and most measurements are carried out at temperatures around $1000\ ^\circ\SIC$ \cite{Udin1949,Kumikov1983}. \textit{Ab initio} calculations can give some impression of the behaviour at lower temperatures \cite{Skriver1992}, but these can hardly stand alone. Extrapolating the linear fit of Ref.~\cite{Udin1949} down to $0~\SIK$ yields surface tension values close to those obtained from \textit{ab initio} calculations in Ref.~\cite{Skriver1992}. This makes it somewhat plausible to apply the model from Ref.~\cite{Udin1949} in the region of interest around $300\ \SIK$. This yields a copper-gas surface energy of $1.92\ \SIJ/\SIm^2$. The contact angle at the copper-water interface is very small \cite{Trevoy1958}, so finding the copper-water surface energy is just a matter of subtracting the surface energy of water from that of copper. The resulting surface energy is $\gamma \approx 1.85\ \SIJ/\SIm^2$, as listed in \tabref{Parameters}.
A great deal of uncertainty is also associated with the value of $k_0$, and the value of $\alpha=\frac12$ is largely a matter of convention.

These choices leave us with three free parameters, which are the bias voltage $V_0$, the electrolyte concentration $c_0$, and the system length $L$.

The solution procedure is as follows: First, the zeroth-order problem is solved for a given set of parameters. Then the first-order problem is solved for a range of wavenumbers $k$. For each $k$ value, the corresponding growth rate $\Gamma$ and perturbation amplitude of the lower electrode, $F_\ell$, are obtained.

In \figref{zero_order_fields_arrow}, the zeroth-order cation concentrations $c_+^0$ and space-charge density $\rho^0$ are shown for $c_0= 10~\SImM$, $L=10~\SImum$ and varying bias voltage $V_0$. It is seen, that when the bias voltage exceeds $V_0 \simeq 12$, local electroneutrality is violated near the cathode. For $V_0=30$ the nonequilibrium space-charge region extends far ($0.04L$) into the electrolyte.

\begin{figure}[!t]
    \includegraphics[width=0.9\columnwidth]{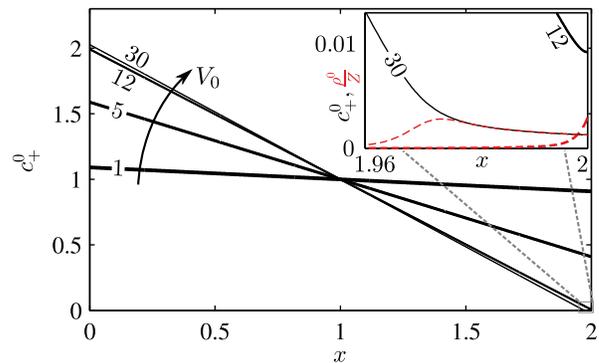}
    \caption{\figlab{zero_order_fields_arrow} (Color online) Zeroth-order cation concentrations $c_+^0$ shown in full (black) lines and zeroth-order charge densities $\rho^0/Z$ shown in dashed (red) lines. The inset shows the fields close to the electrode. In the simulation the parameter values $c_0= 10~\SImM$, $L=10~\SImum$, and $V_0=\{1,5,12,30\}$ were used.}
\end{figure}

\begin{figure}[!t]
    \includegraphics[width=0.9\columnwidth]{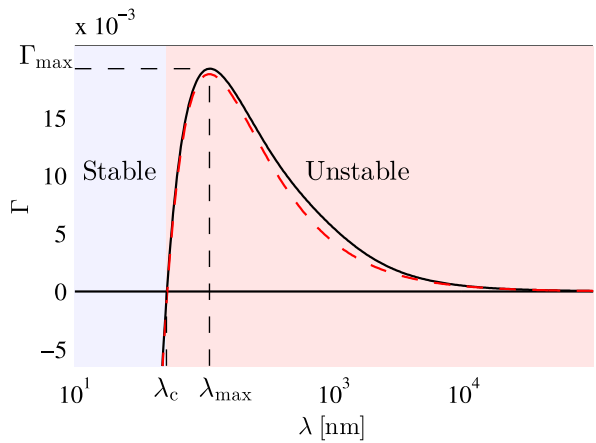}
    \caption{\figlab{Gamma_vs_lambda} (Color online) The growth rate $\Gamma$ plotted versus the perturbation wavelength $\lambda$ for $V_0=30$, $c_0=10~\SImM$, and $L=10~\SImum$. The full (black) line shows the growth rate obtained from numerical simulations, and the dashed (red) line shows the growth rate according to the analytical model \eqref{bGam_analyt}. For perturbation wavelengths smaller than the critical wavelength $\lambda_\mr{c} = 51~\SInm$ the system is stable and for larger wavelengths it is unstable. At the most unstable wavelength $\lambda_\mr{max} = 110 ~\SInm$ the growth rate is $\Gamma_\mr{max} = 0.0193$.  }
\end{figure}

\subsection{Results}

For plotting purposes we introduce the dimensionfull perturbation wavelength $\lambda = 2\pi L/k$. In \figref{Gamma_vs_lambda}, the growth rate $\Gamma$ is plotted versus $\lambda$ for $V_0 =30$, $c_0 = 10~\SImM$, and $L=10~\SImum$. Visible in the figure is a stable region for wavelengths smaller than the critical wavelength $\lambda_\mr{c} = 51~\SInm$, and an unstable region for larger wavelengths. The most unstable wavelength we denote $\lambda_\mr{max}$, and the corresponding growth rate we denote $\Gamma_\mr{max}$.

To enable a more compact representation of the data, we introduce a gray-scale contour plot of the magnitude of $\Gamma$, as illustrated in \figref{Contour_plots_3D}. Here, $\Gamma$ is plotted versus the wavelength $\lambda$ for $V_0=\{5,\ 10,\ 15,\ 20,\ 25,\ 30\}$. The gray scale in the $\lambda$-$V_0$ plane is created by projecting the $\Gamma$ values from the above curves onto the plane. The solid (blue) line in the ($\lambda$, $V_0$)-plane marks the crest of the hill, thus representing the most unstable wavelength for each value of $V_0$.

In \figref{Several_contour_plots}, we make use of the contour plots to show results for twelve sets of $(c_0,L)$-values. In each contour plot, $\Gamma$ is normalized by its maximum value, which is given above each plot. Shown in thick lines are $\lambda_\mr{max}$ in bright (yellow) and $\lambda_\mr{c}$ in black. The corresponding analytical results are shown in dashed (blue) and dotted (green) lines, respectively. The thin black lines show contours, where $\Gamma$ equals $\{0.01, 0.2,0.7\}$ times the maximum value. There is a clear tendency in all of the panels that the growth rate $\Gamma$ increases rapidly with $V_0$, and the most unstable wavelength decreases as $V_0$ increases. Across the panels, the maximum growth rate is seen to increase for increasing $c_0$ and increasing $L$. Also, the most unstable wavelength $\lambda_\mr{max}$ and the critical wavelength $\lambda_\mr{c}$ become smaller as $c_0$ increases and as $L$ decreases.

A common feature seen in all of the panels, is the kink in the $V_0$-versus-$\lambda_\mr{max}$ and $V_0$-versus-$\lambda_\mr{c}$ lines. At this kink, the slope of the lines changes markedly. The kink is located at the voltage, where the current reaches the limiting current, and it thus signifies that there is a qualitatively different behavior for over- and underlimiting current. This qualitative difference between the two regimes is in accordance with the analytical models. We also see that the kink voltage changes with $c_0$ and $L$. Specifically, it increases with $c_0$ and decreases with $L$. The main reason for this behavior is easily understood with reference to the zeroth-order Butler-Volmer reaction expression~\eqrefNoeq{Zero_order_reaction}. Setting the current in the system to the limiting current $J_+^0=1$, the reaction rates at the electrodes become
 \begin{align}
 \vec{e}_x \cdot \vn_p =- K_0 \left [c_+ e^{\alpha Z(\phi + V_p)} - e^{-(1-\alpha)Z(\phi+V_p)}\right ].
 \end{align}
At the cathode, the first term in the bracket dominates, and at the anode the other. Therefore, both potential drops over the electrode interfaces scale as
\begin{align}
\Delta V \sim -\ln(K_0) = \ln\left (\frac{2 D_+ c_0}{L}\right )  -\ln(k_0),
\end{align}
which increases monotonically with increasing $c_0/L$. As a consequence, the total potential drop at the limiting current also increases with increasing $c_0/L$, just as observed in \figref{Several_contour_plots}.

\begin{figure}[!t]
    \includegraphics[width=0.9\columnwidth]{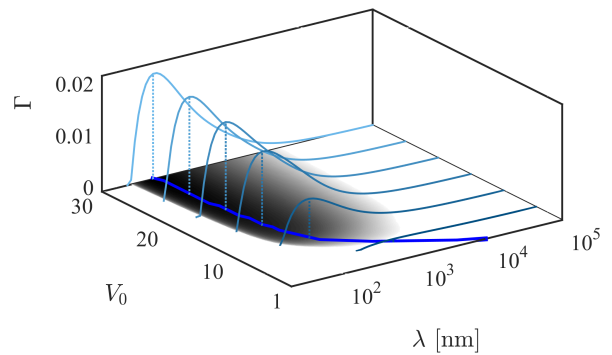}
    \caption{\figlab{Contour_plots_3D} (Color online) The growth rate $\Gamma$ plotted versus the perturbation wavelength $\lambda$ and voltage $V_0$ for $c_0=10~\SImM$, and $L=10~\SImum$. The (cyan) space curves are plots of $\Gamma$ versus $\lambda$ for $V_0~=~\{5,\ 10,\ 15,\ 20,\ 25,\ 30\}$. The shade of the in plane contour plot is based on the logarithm of $\Gamma$, which is why there are no contours in the low $\lambda$ limit where $\Gamma$ is negative. The thick (blue) in plane line marks the crest of the hill, i.e. it marks the most unstable wavelength for each value of $V_0$.}
\end{figure}

\begin{figure*}[!t]
    \includegraphics[width=17 cm]{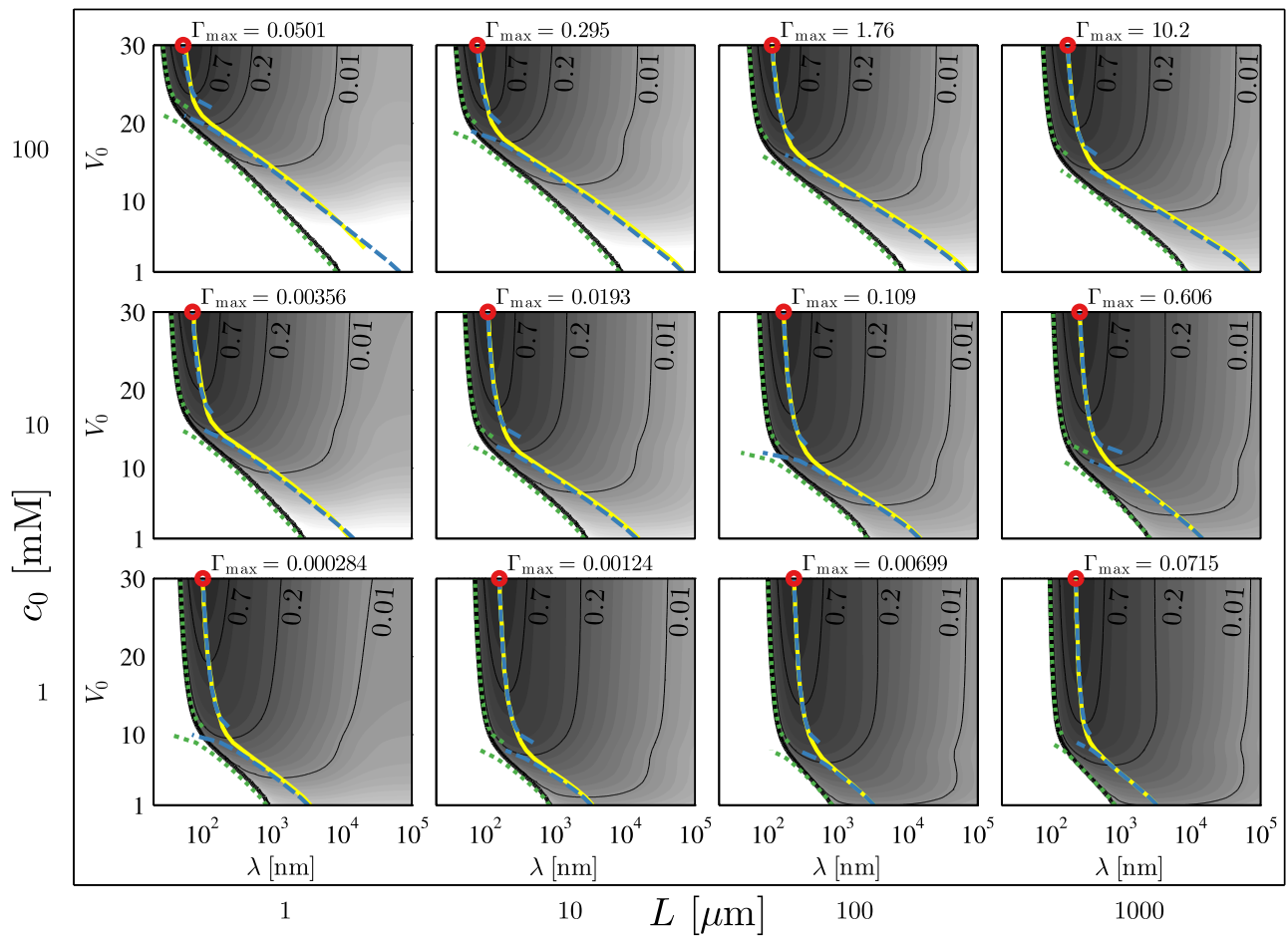}
    \caption{\figlab{Several_contour_plots} (Color online) Contour plots of $\Gamma$ plotted versus wavelength $\lambda$ and voltage $V_0$ for $c_0=\{1~\SImM ,10~\SImM ,100~\SImM \}$ and $L~=~\{1~\SImum,10~\SImum,100~\SImum, 1~\SImm\}$. In each plot, $\Gamma$ is normalized by its maximum value, and the contours are logarithmically spaced. The maximum value $\Gamma_\mr{max}$ of $\Gamma$ is given on top of each plot, and the point where the maximum value is attained is indicated with a dark (red) circle. The three thin black lines in each plot indicate contours where $\Gamma$ equals $0.01$, $0.2$, and $0.7$ times $\Gamma_\mr{max}$. The thick bright (yellow) line marks $\lambda_\mr{max}$ for each value of $V_0$, and the dashed (blue) lines mark the two corresponding analytical limits. The thick black line marks $\lambda_\mr{c}$ for each value of $V_0$, and the dotted (green) lines mark the two corresponding analytical limits.}
\end{figure*}

In addition to the instability growth rate $\Gamma$, which gives a time scale for the development of instabilities, it is useful to have a measure for the characteristic instability length scale. For instance, we would like to estimate the thickness of the deposited layer, when instabilities start to develop. We define this instability length scale as the product of the zeroth-order growth rate \eqref{zero_order_electrode_growth} and the instability time scale at the most unstable wavelength
 \beq{LGam1}
 L_\Gamma =  L \frac{a^3c_0 J_+^0}{\Gamma_\mr{max}},
 \eeq
where the pre-factor $L$ ensures a dimensionfull expression. In \figref{Growth_length}, we plot the instability length $L_\Gamma$ versus applied voltage $V_0$ for $L=100~\SImum$ and varying $c_0$. The most unstable wavelength $\lambda_\mr{max}$ is also plotted in the same figure (dashed lines). It is seen that $L_\Gamma$ decreases as $V_0$ increases, but for small voltages $L_\Gamma$ is largest for high concentrations, while the opposite is true for high voltages. The reason for this reversal is that the interfacial voltage drops are largest for large $c_0$. At small voltages the bulk driving force in the systems with large $c_0$ is therefore small, and this causes the system to be less unstable than the low $c_0$ systems. We also see that $\lambda_\mr{max}$ scales in the same way as $L_\Gamma$. While the reason for this is not immediately obvious, it is seen to follow from the analytical expressions. Inserting \eqref{bGam_analyt} in \eqref{LGam1} yields
\begin{align} \eqlab{LGam2}
L_\Gamma =  \frac{\lambda_\mr{max}}{2\pi} \frac{\xi + \frac{2\pi L}{\lambda_\mr{max}}}{\xi-\bgam \big( \frac{2\pi L}{\lambda_\mr{max}}\big)^2},
\end{align}
which confirms the approximate scaling between $L_\Gamma$ and $\lambda_\mr{max}$. The connection between $L_\Gamma$ and $\lambda_\mr{max}$ implies that $\lambda_\mr{max}$ sets the scale, not only for the variations in the horizontal direction, but also for variations in the vertical direction. We might therefore expect that the ramified electrodeposits, emerging at much longer times than $\Gamma_\mr{max}^{-1}$, have a universal length scale roughly set by $\lambda_\mr{max}$.

\begin{figure}[!t]
    \includegraphics[width=0.9\columnwidth]{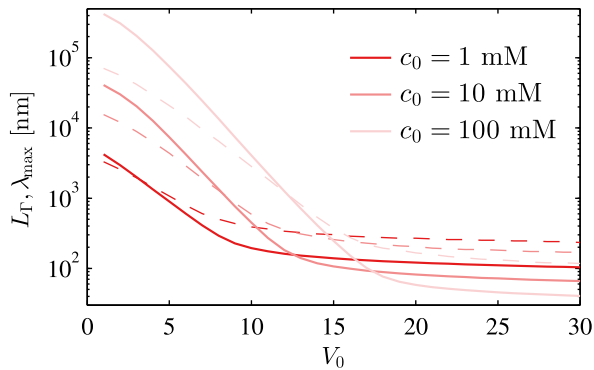}
    \caption{\figlab{Growth_length} (Color online) The instability length scale $L_\Gamma$ (full line) and most unstable wavelength $\lambda_\mr{max}$ (dashed line) plotted versus bias voltage $V_0$. The concentration varies between the values $c_0 = \{1~\SImM,10~\SImM,100~\SImM\}$ and the length $L=100~\SImum$ was used. }
\end{figure}

\section{Discussion}

The main feature, which sets our work apart from previous stability analyses of electrodeposition, is the inclusion of the overlimiting regime. Presumably, this regime has so far been avoided due to the non-linearities arising at overlimiting current, which necessitate a more complicated treatment. However, the overlimiting regime is highly relevant for ramified growth problems \cite{Han2014,Gonzalez2008}. As seen in \figref{Several_contour_plots}, the instability growth rate increases markedly in the overlimiting regime, and there is also a change in qualitative behavior between the two regimes. Of course, the conclusions we reach, based on our model, are only strictly valid for planar electrodes. It does, however, seem reasonable to expect that the most unstable wavelength $\lambda_\mr{max}$ is comparable to the characteristic dimensions encountered in a ramified growth experiment. Our analysis can thus be used to rationalize experimental results. In this regard, our analytical models are particularly useful, since they allow for easy computation of the key parameters for other systems than the one treated here.

Perhaps the most important application of the stability analysis, is as a means of validating more elaborate numerical models of ramified growth. A model of ramified growth must necessarily deal with a moving interface and this, as well as other complications, make for highly complex numerical models. To validate such models it is very useful to have a comparatively simple model, like the present one, to benchmark against in the relevant limit. Indeed, this was what originally motivated us to treat the stability problem.

An obvious shortcoming of the given analysis, is the restriction to a steady-state zeroth-order solution. The principal reason for this choice is that it makes for a simpler problem. Furthermore, the numerical ramified growth model, to which we wish to compare our model, is at present also restricted to quasi-steady state. In time, we wish to extend both models to the fully transient regime.

There is, however, some physical justification for making the steady-state assumption. As seen in \figref{Several_contour_plots}, the growth rate $\Gamma$ is considerably smaller than unity in a large part of the investigated parameter space. The time it takes the system to reach steady state is given by the diffusive time, which in our normalization has the value one. Thus, as long as $\Gamma$ is much smaller than unity, the system reaches steady state long before any instabilities build up. In this case it is therefore justified to assume steady state.
It should be noted that in this argument we make the reasonable assumption that the true growth rate in the transient regime does not significantly exceed the steady-state value.

To model the reaction rate at the electrodes we use the standard Butler-Volmer expression, which stands at the core of much electrochemistry. The Butler-Volmer model is, however, largely a phenomenological model, that does not necessarily apply in all regimes \cite{Bazant2013}. As applied here, the model does not distinguish between the potential drop over the Debye layer and the potential drop in the narrow interface region. This does not seem quite right, but it could be fixed with relative ease by implementing the Frumkin correction to the Butler-Volmer model. Nevertheless, to be in accordance with most previous studies, and because the Frumkin correction would introduce another unknown parameter, we have chosen not to make this correction.

\section{Conclusion}

We have successfully solved the stability problem in the under- and overlimiting regime for the case of a copper sulfate solution trapped between two copper electrodes. In addition to the numerical solution of this particular problem, we have derived analytical solutions valid in either the overlimiting or the underlimiting limit. The behavior in the overlimiting regime differs qualitatively from the behavior in the underlimiting regime, and we find that the electrode becomes increasingly unstable as the current is increased above the limiting current. The stability analysis, and in particular the analytical limits, are valuable both for rationalizing experimental results and for validating more elaborate numerical models of ramified growth.

\appendix

\section{The electroneutral limit}
\chaplab{App_Lin}

In the limit where the electrolyte is locally electroneutral and the time derivatives in the first-order transport problem are negligible, analytical solutions to the problem can be obtained. Setting the point of zero electrostatic potential at $x=1$, it is easily found that
 \begin{align}
 c =c_+ =c_- = e^{Z\phi}.
 \end{align}
It follows that $c = e^{Z(\phi^0+\phi^1)} \approx e^{Z\phi^0} + e^{Z\phi^0}Z\phi^1$, and thus
 \beq{LEN_first_and_zero_order_fields}
 c^0 = e^{Z\phi^0} \quad \text{ and } \quad c^1= e^{Z\phi^0}Z\phi^1.
 \eeq
Solving the zeroth-order problem yields
\begin{align}
c^0 = 1-J^0 (x-1), \quad Z\phi^0 = \ln(1-J^0 (x-1)).
\end{align}
Using the electroneutrality assumption in \eqref{Star_bulk_eq} we find
\begin{align}
0 = \pp_x^2 c^* - k^2 c^*.
\end{align}
This equation has two solutions, but as long as the perturbation wavelength is considerably smaller than the electrode spacing, the solution which increases with $x$ is dominant
\begin{align}
c^* \approx C e^{k(x-2)},
\end{align}
where $C$ is a constant to be determined.
From \eqref{LEN_first_and_zero_order_fields} we then find
\begin{align}
\phi^* = \frac{1}{Z} \frac{c^*}{c^0} = \frac{C}{Z} \frac{e^{k(x-2)}}{1-J^0(x-1)}.
\end{align}
At the upper electrode, $x=2$, the first-order reaction rate is (we set $F_u=1$)
 \begin{align}
 &\frac{R_u^*}{K_0} =
 \nn \\
 &e^{\alpha Z(\phi^0+V_u)}
 \bigg[ c^*+ \pp_x c^0  +c_+^0\Big(\bgam k^2+\alpha Z\big[\phi^*+\pp_x \phi^0\big]\Big) \bigg]
 \nn \\
 &-e^{ - (1-\alpha)Z(\phi^0+V_u) }\bigg[\bgam k^2  -   (1-\alpha)Z\big[\phi^*+\pp_x \phi^0\big] \bigg].
 \end{align}
We can simplify this expression by using
\begin{align}
e^{ - (1-\alpha)Z(\phi^0+V_u) } = c^0e^{\alpha Z(\phi^0+V_u)} - \frac{R_u^0}{K_0},
\end{align}
from the zeroth-order reaction expression. The first-order reaction expression then becomes
\begin{align}
\frac{R_u^*}{K_0} &=  e^{\alpha Z(\phi^0+V_u)}\bigg[ c^*+ \pp_x c^0  +c_+^0 Z\big(\phi^*+\pp_x \phi^0 \big) \bigg] \nn \\
&\quad +\frac{R_u^0}{K_0}\bigg[\bgam k^2  -   (1-\alpha)Z\big(\phi^*+\pp_x \phi^0 \big) \bigg]. \eqlab{Simpler_first_order_reaction}
\end{align}
Evaluating the fields at $x=2$, this expression becomes
 \begin{align}
 \frac{R_u^*}{K_0} &= 2(C-J^0) e^{\alpha Z(\phi^0+V_u)}
 \nn \\
 &\quad +\frac{J^0}{K_0}\bigg[\bgam k^2  -   (1-\alpha)\frac{C-J^0}{1-J^0} \bigg].
 \end{align}
The current into the upper electrode is $J^* = -\pp_xc^* = -kC$, meaning that
 \begin{align}
 -kC = R_u^* &= 2(C-J^0)K_0e^{\alpha Z(\phi^0+V_u)}
 \nn \\
 &\quad +J^0\bgam k^2  -   (1-\alpha)J_0\frac{C-J^0}{1-J^0} ,
 \end{align}
and solving for $C$, we obtain
\begin{align}
C&=J^0\frac{-2K_0e^{\alpha Z(\phi^0+V_u)} +\bgam k^2  +   (1-\alpha)\frac{J^0}{1-J^0}}{-2K_0e^{\alpha Z(\phi^0+V_u)}-k+(1-\alpha)\frac{J^0}{1-J^0}}.
\end{align}
The growth rate can be expressed as
\begin{align}
\Gamma = -a^3c_0J^* = a^3c_0kC,
\end{align}
so we have
\begin{align}
\Gamma = a^3c_0kJ^0\frac{2K_0e^{\alpha Z(\phi^0+V_u)} -\bgam k^2  -  (1-\alpha)\frac{J^0}{1-J^0}}{2K_0e^{\alpha Z(\phi^0+V_u)}+k-(1-\alpha)\frac{J^0}{1-J^0}}
\end{align}
For the special case where $\alpha=\frac{1}{2}$ we find from the zeroth-order current that
\begin{align}
2 K_0 e^{\alpha Z(\phi^0+V_u)} = \frac{J^0}{c^0}\left [1+\sqrt{1+4\left (\frac{K_0}{J^0}\right )^2c^0}\right ]. \eqlab{When_alpha_is_a_half}
\end{align}
Inserting this yields
 \begin{align}  \eqlab{bGam_Lin}
 \Gamma = a^3c_0kJ^0\frac{\frac{J^0}{1-J^0}\left [\alpha +\sqrt{1+4\left (\frac{K_0}{J^0}\right )^2(1-J^0)}\right ] -\bgam k^2  }{\frac{J^0}{1-J^0}\left [\alpha +\sqrt{1+4\left (\frac{K_0}{J^0}\right )^2(1-J^0)}\right ]+k},
 \end{align}
an expression which accurately replicates the numerical results at low voltages and low $\blamD$.

To test whether the time derivatives in the first-order problem really are negligible, we compare the time derivative term $2 \Gamma c_+^*$ with the transverse diffusion term $k^2 c_+^*$. Since \eqref{bGam_Lin} implies $\Gamma \leq a^3c_0kJ^0$, our assumption is justified if
 \begin{align}
 2a^3c_0J^0 \ll k.
 \end{align}
Consequently, because $a^3c_0\ll 1$ for dilute systems and $J^0$ is of order unity, it is justified to neglect the time derivative, unless the perturbation wavelength is much larger than the electrode spacing.

The critical wavenumber $k_c$ is found by setting the nominator in \eqref{bGam_Lin} equal to zero,
 \begin{align}  \eqlab{kc}
 k_\mr{c} = \sqrt{ \frac{\xi}{\bgam}},
 \end{align}
where we have introduced the parameter
 \begin{align}  \eqlab{xi_lin}
 \xi =\frac{J^0}{1-J^0}\left [\alpha +\sqrt{1+4\left (\frac{K_0}{J^0}\right )^2(1-J^0)}\right ].
 \end{align}
To find the wavenumber $\kmax$, at which $\Gamma$ attains its maximum $\Gamma_\mr{max}$, we set the derivative of $\Gamma$ equal to zero and solve for $k$,
 \begin{align}  \eqlab{kmax}
 \kmax &= \frac{\xi}{2}\bigg [ \left (\frac{2-\xi\bgam+2\sqrt{1-\xi\bgam}}{\xi\bgam}\right )^{1/3}
 \nn \\
 &\quad +\left (\frac{2-\xi\bgam+2\sqrt{1-\xi\bgam}}{\xi\bgam}\right )^{-1/3}  -1 \bigg],
 \end{align}
with the asymptotic solutions,
\begin{align}
\kmax \approx \left\{ \begin{array}{lcl}
\left (\frac{\xi}{3\bgam}\right )^{1/2}, & \text{for} & \bgam\xi \gg 1,     \\
\left (\frac{\xi^2}{2\bgam}\right )^{1/3}-\frac{\xi}{2}, &  \text{for} & \bgam\xi \ll 1.
\end{array}  \right.
\end{align}

\section{The strongly non-linear limit}
\chaplab{App_Nonlin}

In the limit where the driving force is very large, some of the terms in \eqsref{Star_bulk_eq}{Star_Poisson} become dominant, which makes an analytical solution of the problem possible.

If the system is strongly driven, the field gradients are large close to the upper electrode, and this makes the electrode surface much more unstable. It follows that a larger $k$ value is needed for the surface tension to stabilize the system, so the most unstable value of $k$ will be larger than for less driven systems. In the strongly driven limit, we might therefore expect that \eqref{Star_Poisson} largely is a balance between $\pp_x^2\phi^*$ and $k^2 \phi^*$ in the region of interest. This leads us to making the ansatz
\begin{align} \eqlab{phi_ansatz}
\phi^* = \Phi e^{k(x-2)},
\end{align}
where $\Phi$ is a constant. We now consider \eqref{Star_bulk_eq} for the cation concentration, neglecting the left hand side
\begin{align}
0 &= - \pp_x \bigg\{ - \pp_x c_+^*- Z c_+^*\pp_x\phi^0- Z c_+^0\pp_x \phi^*  \bigg\}\nn  \\
&\quad - k^2 (c_+^* + Zc_+^0\phi^*).
\end{align}
We assume that the terms $\pp_x c_+^*$ and $Z c_+^*\pp_x\phi^0$ are negligible compared to $Z c_+^0\pp_x \phi^*$ and insert the ansatz \eqref{phi_ansatz}
\begin{align}
0 &\approx   Z \pp_x c_+^0k\phi^* + Z c_+^0k^2\phi^* - k^2 (c_+^* + Zc_+^0\phi^*) \\
&\approx Z \pp_x c_+^0k\phi^*- k^2 c_+^*,
\end{align}
implying that
 \begin{align}
 c_+^* \approx \frac{Z}{k} \pp_x c_+^0\phi^*.
 \end{align}
To test the assumptions leading to this result, we need expressions for $ c_+^0$, $\pp_x c_+^0$ and $\pp_x \phi^0$. From Ref.~\cite{Nielsen2014} we have such expressions, and in the extended space-charge region (ESC) they take the simple forms
\begin{align}
c_+^0(x) &\approx \sqrt{2}\frac{\blamD}{Z} \sqrt{J^0} \left [x-1-\frac{1}{J^0}\right ]^{-1/2}, \\
\pp_x c_+^0(x) &\approx -\frac{\sqrt{2}}{2}\frac{\blamD}{Z} \sqrt{J^0} \left [x-1-\frac{1}{J^0}\right ]^{-3/2},\\
\pp_x \phi^0(x) & \approx - \frac{\sqrt{2}}{\blamD} \sqrt{J^0} \left [x-1-\frac{1}{J^0}\right ]^{1/2}.
\end{align}
The width of the ESC is given as $L_\mr{ESC} = 1-1/J^0$, so in the region close to the electrode, compared to the width of the ESC, the fields can be written as
\begin{align}
c_+^0(x) &\approx \sqrt{2}\frac{\blamD}{Z} \sqrt{J^0} L_\mr{ESC}^{-1/2}, \\
\pp_x c_+^0(x) &\approx -\frac{c_+^0}{2 L_\mr{ESC}},\\
\pp_x \phi^0(x) & \approx - \frac{Z c_+^0}{\blamDsqr}L_\mr{ESC}.
\end{align}
Evaluating $\pp_x c_+^*$ we find
\begin{align}
\pp_x c_+^* \approx \frac{Z}{k} \frac{3c_+^0}{4L_\mr{ESC}^2 } \phi^* - Z \frac{c_+^0}{2L_\mr{ESC}} \phi^*,
\end{align}
which is seen to be much smaller than $Z c_+^0 \pp_x \phi^*$ if
\begin{align}
2k \gg \frac{1}{L_\mr{ESC}},
\end{align}
that is, if the perturbation wavelength satisfies
\begin{align} \eqlab{blam_ll_LESC}
\blam \ll 4 \pi L_\mr{ESC}.
\end{align}
Similarly, we find that $Z c_+^*\pp_x\phi^0$ is much smaller than $Z c_+^0\pp_x \phi^*$ if
\begin{align}   \eqlab{blam_ll_blamD}
\blam^2 \ll \frac{8\pi^2}{Z^2} \frac{\blamDsqr}{c_+^0(2)}.
\end{align}
Finally, the ansatz \eqref{phi_ansatz} is justified if $2\blamDsqr k^2 \phi^* \gg Z c_+^*$, which is equivalent to
\begin{align}
\blam^3 \ll \frac{32 \pi^3}{Z^2}\frac{\blamDsqr}{c_+^0(2)}L_\mr{ESC}.
\end{align}
This last requirement is seen to follow if the two first requirements \eqsref{blam_ll_LESC}{blam_ll_blamD} are fulfilled.

In the strongly driven regime, where \eqsref{blam_ll_LESC}{blam_ll_blamD} are satisfied, the first-order current is approximately
\begin{align}
2J^*_+ \approx -Z c_+^0\pp_x \phi^* = -Z kc_+^0 \Phi,
\end{align}
at the upper electrode. The zeroth-order diffusive contribution is also very small at the upper electrode, meaning that we can simplify \eqref{Simpler_first_order_reaction}
\begin{align}
R_u^* &\approx   K_0e^{\alpha Z(\phi^0+V_u)}\bigg[ c_+^0 Z\big(\phi^*+\pp_x \phi^0 \big) \bigg] \nn \\
&\quad +R_u^0\bigg[\bgam k^2  -   (1-\alpha)Z\big(\phi^*+\pp_x \phi^0 \big) \bigg] \\
&\approx  K_0e^{\alpha Z(\phi^0+V_u)}\bigg[ c_+^0 Z\Phi-2J^0  \bigg] \nn \\
&\quad +R_u^0\bigg[\bgam k^2  -   (1-\alpha)\bigg(Z\Phi-\frac{2J^0}{c_+^0} \bigg) \bigg].
\end{align}
Inserting $R_u^* = J^*_+ \approx -\frac{1}{2}Z kc_+^0 \Phi$ we find
\begin{align}
\frac{Z}{2} kc_+^0 \Phi = kJ^0\frac{2K_0e^{\alpha Z(\phi^0+V_u)} -\bgam k^2  -  (1-\alpha)\frac{2J^0}{c_+^0}}{2K_0e^{\alpha Z(\phi^0+V_u)}+k-(1-\alpha)\frac{2J^0}{c_+^0}},
\end{align}
and since $\Gamma = - a^3c_0J^*$  \\[20mm]
\begin{align}
\Gamma &= a^3c_0kJ^0\frac{2K_0e^{\alpha Z(\phi^0+V_u)} -\bgam k^2  -  (1-\alpha)\frac{2J^0}{c_+^0}}{2K_0e^{\alpha Z(\phi^0+V_u)}+k-(1-\alpha)\frac{2J^0}{c_+^0}} \\
&= a^3c_0kJ^0\frac{\frac{J^0}{c_+^0}\left [2\alpha-1 +\sqrt{1+4\left (\frac{K_0}{J^0}\right )^2c_+^0}\right ] -\bgam k^2  }{\frac{J^0}{c_+^0}\left [2\alpha-1 +\sqrt{1+4\left (\frac{K_0}{J^0}\right )^2c_+^0}\right ]+k},
\end{align}
where the second expression assumes $\alpha=\frac{1}{2}$, so that \eqref{When_alpha_is_a_half} is valid. Like in the electroneutral limit neglecting the time derivative in the first-order problem is justified, unless the perturbation wavelength is much larger than the electrode spacing. The expressions \eqrefNoeq{kc} and \eqrefNoeq{kmax} are also valid for the strongly nonlinear limit, if instead of \eqref{xi_lin} we use
\begin{align} \eqlab{xi_nonlin}
\xi =\frac{J^0}{c_+^0}\left [2\alpha-1 +\sqrt{1+4\left (\frac{K_0}{J^0}\right )^2c_+^0}\right ].
\end{align}

%

\end{document}